\documentclass[twocolumn,floatfix,preprintnumbers,superscriptaddress]{revtex4-1}

\usepackage[utf8]{inputenc}
\usepackage[colorlinks=true,citecolor=blue,linkcolor=blue]{hyperref}
\usepackage[normalem]{ulem}
\usepackage{url}
\usepackage{graphicx,wrapfig,float,slashed,cancel}
\usepackage{amsmath,amssymb,epsfig,graphicx,xcolor,stmaryrd}
\usepackage{bm}
\usepackage{enumitem}
\usepackage{subfigure}
\usepackage{amsfonts}
\usepackage{amssymb}
\usepackage{dsfont}
\usepackage{multirow}
\usepackage{appendix}
\usepackage{slashed}
\usepackage[active]{srcltx}
\usepackage{psfrag}
\usepackage{multirow}
\input epsf
\definecolor{darkblue}{RGB}{1, 90, 173}

\begin{document}   

\title{Properties of kaon at non-zero temperature and baryon chemical potential}
\date{\today}

\author{G. Bozk{\i}r }
\affiliation{Department of Basic Sciences,  Army NCO Vocational HE School, National Defense University, Alt{\i}eyl\"{u}l, 
10185 Bal{\i}kesir,Turkey}
\author{A. T{\"u}rkan}
\affiliation{Department~of Natural~and~Mathematical~Sciences, \"Ozye\v{g}in University,  \c{C}ekmek\"{o}y, 
34794 Istanbul, Turkey}
\author{K.~Azizi}
\thanks{Corresponding Author}
\affiliation{Department of Physics, University of Tehran,  North Karegar Avenue, Tehran,
	14395-547 Iran}
\affiliation{Department of Physics, Do\v{g}u\c{s} University, Dudullu-\"{U}mraniye, 34775
Istanbul, Turkey}

\begin{abstract}
We investigate the spectroscopic properties of the strange particle kaon in the framework of hot and dense QCD. To this end, first, we find the perturbative spectral density, which is connected with both the  temperature $T$ and the  baryon chemical potential $\mu_{B}$. We include the non-perturbative operators as functions of temperature and baryon chemical potential up to mass dimension five. We perform the calculations in momentum space and use the quark propagator in the hot and dense medium. The numerical results at non-zero temperature and baryon chemical potential demonstrate that the mass of the particle rises considerably by increasing the baryon chemical potential at a fixed temperature (for both the zero and non-zero temperatures) up to approximately $\mu_{B}=0.4$ GeV. After this point, it starts to fall by increasing the baryon chemical potential and it apparently vanishes  at $\mu_{B}=(1.03-1.15)$ GeV for finite temperatures: The point of apparent vanishing  moves to lower baryon chemical potentials by increasing the temperature. At zero temperature, the mass reaches to roughly a fixed value at higher baryon chemical potentials. On the other hand, the decay constant decreases considerably with respect to baryon chemical potential up to roughly $\mu_{B}=0.4$ GeV, but after this point, it starts to increase in terms of the baryon chemical potential at finite temperatures. At $T=0$, the decay constant reaches to a fixed value at higher chemical potentials, as well. Regarding the dependence on the temperature we observe that, at fixed values of baryon chemical potentials, the mass and decay constant remain roughly unchanged up to $T=50$ MeV and  $T=70$ MeV  respectively, but after these points, the mass starts to fall and the decay constant starts to rise up to a critical temperature  $T=155$ MeV, considerably.  It is also seen that the obtained results for the mass and decay constant at $T=\mu_{B}=0$ are in good consistency with the existing experimental data. The observations are consistent with the QCD phase diagram in the $T-\mu_{B}$ plane.

\end{abstract}

\maketitle
  
%%%%%%%%%%%%%%%%%%%%%%%%%%%%%%%%%%%%%%%%%%%%%%%%%%%%%%%%%%%%%%%%%%%%%%%%%%%%

\section{Introduction}  \label{sec:intro}

\begin{figure}[ht]
\begin{center}
\includegraphics[width=8.5cm]{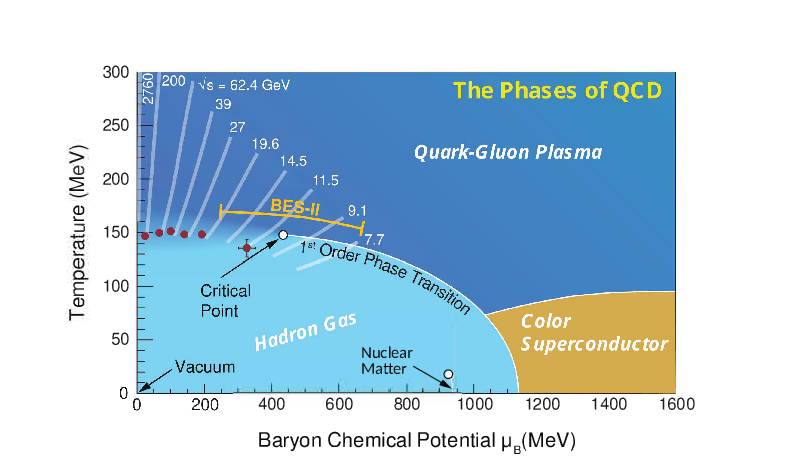}
\end{center}
\caption{QCD phase diagram in $T-\mu_{B}$ plane \cite{Aprahamian}.} \label{fig1}
\end{figure}

In particle physics, the quantum field  theory of strongly interacting subatomic particles is called Quantum Chromodynamics (QCD). The investigation of possible crossover and  phase transition of hadronic gas under extreme conditions such as the  high temperature and baryon chemical potential is one of the most important research topics in QCD. It provides insight into a better understanding of the hadronic matter and quark-gluon plasma (QGP) as a possible new state of matter. The crossover, first order and second order phase transitions may occur in relativistic heavy-ion collision experiments, the center of neutron stars and the early universe. A QCD phase diagram is shown in Fig.  (\ref{fig1}) \cite{Aprahamian}. The QCD phase diagram was first pictured by Cabibbo and Parisi in 1975 \cite{Cabibbo}. They showed that the baryonic density decreases exponentially with an increase in temperature such that at high enough temperatures quarks move  relatively freely interpreted as a situation connected with the existence of a different phase. In Ref. \cite{Asakawa}, the presence of the end point in the first order phase transition was defined by using Nambu-Jona-Lasinio (NJL) model that then it was called as critical end point (CEP) in the QCD phase diagram.

The main purpose of studies on the QCD phase diagram is to determine the location of crossover (at high temperature-zero baryon chemical potential), the curve of first order phase transition, and superconductivity in $T-\mu_{B}$ plane. For the crossover phase transition, the (pseudo)critical temperature  $T_{c}$ at which quarks and gluons deconfined has been explicitly given by Relativistic Heavy Ion Collider (RHIC) at Brookhaven National Laboratory (BNL) \cite{Arsene,Back,Adams,Adcox,Gyulassy} and Large Hadron Collider (LHC) at CERN \cite{Singh}. Experimental signals of CEP have been proposed in Super Proton Synchroton (SPS) at CERN \cite{Gazdzicki1,Gazdzicki2} and RHIC at BNL \cite{Stephanov1,Stephanov2}. In the next years, experimental studies for the higher baryon chemical potentials will be conducted at the Facility for the Compressed Baryonic Matter Experiment (CBM) at Antiproton and Ion Research (FAIR) in Germany, the Nuclotron-based Ion Collider fAcility (NICA) at the Joint Institute for Nuclear Research (JINR) in Russia and the J-PARC heavy ion project (J-PARC-HI) at the Japan Proton Accelerator Research Complex (J-PARC). Also, at very large baryon chemical potentials and the temperature order $10$-$50$ MeV, the color superconducting phase of quark matter may arise in the limit of weak coupling \cite{Dirk}. 

In the theoretical side, the first principle lattice QCD calculations account for smooth crossover occured at $\mu_{B}=0$ and $154MeV\leq T \leq158.6MeV$ \cite{Cheng,Aoki1,Aoki2,Bazavov1,Borsanyi1,Borsanyi2}, but it has come up short for $\mu_{B}>0$ due to the sign problem \cite{Karsch,Muroya,Bernard,Aoki3,Bazavov2,Bedaque}. To overcome of this sign problem in lattice QCD simmulations, several methods such as Taylor expansion in $\mu_{B}$ \cite{Allton1,Allton2,Gavai,Basak,Kaczmarek}, imaginary $\mu_{B}$ \cite{Fodor1,Fodor2,Fodor3,Forcrand1,DElia1,Wu,DElia2,Conradi,Forcrand2,DElia3,Moscicki} and the canonical ensemble \cite{Alexandru,Kratochvila,Ejiri} have been developed at small baryon chemical potential and finite temperature. Also, to investigate QCD phase transitions at small baryon chemical potential and zero temperature, low-energy effective models like NJL model \cite{Casalbuoni,He1,Xia,He2}, chiral perturbation theory \cite{Karsch,Kogut,Brandt,Begun,Son,Loewe}, quark-meson model \cite{Schaefer,Tetradis} and random matrix model \cite{Klein,Vanderheyden} have been used in theoretical calculations.  There is another powerful non-perturbative method called QCD sum rule to investigate not only  the properties of hadrons in vacuum and under extreme conditions but also to search for the possible signs of hadronic phase transition. This method that was first proposed for vacuum \cite{Shifman} and then extended to $T\neq0$ \cite{Bochkarev}, has been used to investigate the hadrons at $T>0$, $\mu_{B}=0$ \cite{Dominguez1,Dominguez2,Dominguez3,Dominguez4,Dominguez5,Aarts1,Aarts2,Mallik1,Mallik2,
	Mallik3,Veliev1,Veliev2,Veliev3,AziziTurkan,Azizi1,Azizi2,Azizi3,Azizi4,Azizi6} as well as  $\mu_{B}>0$, $T=0$ (in nuclear medium) (see for instance \cite{Er1,Er2,Er3,Er4,Er5,Er6,Er14}). Comparatively, few studies have been devoted to the investigation of hadronic properties  at  $T>0$, $\mu_{B}>0$ region of the QCD phase diagram   in the literature \cite{Ayala1,Ayala2,Serna,Tawfik,Abu-Shady}. In Ref. \cite{Ayala1,Ayala2}, to study the QCD phase diagram at $T\neq 0$ and $\mu_{B}\neq 0$,  the authors have obtained the light-quark condensate $\langle \overline{\psi} \psi \rangle \left( T,\mu_{B}\right) $ using QCD Finite Energy Sum Rules (FESR). They also  computed the perturbative QCD threshold $s_{0} \left( T,\mu_{B}\right)  $ from FESR using  the light-quark condensate at finite temperature and density as an  input, determined in the Schwinger-Dyson
equations (SDE) framework. The authors  investigated heat capacities for both the quark condensate and the perturbative QCD threshold $s_{0}$ as a function of $T$ for $\mu_{B}=0$ and $\mu_{B}=300$ MeV. They suggested that the position of the critical end point to be $\mu_{B}\geq 300$ MeV and $T_{c}\leq 185$ MeV. In Ref. \cite{Serna}, the masses of the pseudoscalar mesons $\pi^{+}$, $K^{0}$ and $D^{+}$ have been calculated at finite $T$ and $\mu_{B}$ based on symmetry-preserving Dyson-Schwinger equation treatment of a vector-vector four quark contact interaction. It has been found that obtained results are in qualitative agreement with lattice QCD data and QCD sum rules calculations. In Ref. \cite{Tawfik}, the quark–antiquark condensates for light and strange quark flavors at finite temperatures and chemical potentials have been calculated using a hadron resonance gas model. In Ref. \cite{Abu-Shady}, the nucleon properties at finite temperature and chemical potential are investigated in the framework of the logarithmic quark sigma model. It has been shown that the nucleon's hedgehog mass and the magnetic moment increase with increasing the temperature and chemical potential, while the pion-nucleon coupling constant decreases in these conditions.

Investigation of particles made of strange quark (like kaons) in hot and dense medium is of great importance as their production plays a key role in describing the hot and dense fireball created during the collisions. The strange quark has a mass compatible with  the characteristic temperatures of the nuclear fireball. The kaons  together with pions are  important  for finding one of the predicted signals of the formation of quark-gluon plasma, so called the strangeness enhancement.
For a review on the status of the experimental and theoretical developments in the field of strangeness in nuclei and neutron stars see for instance \cite{Tolos1}. In this study the authors widely discuss the  kaons, antikaons, $\phi$ and hyperons and their interactions with nucleons and nuclear matter.
 In \cite{Miyamura},  the kaon and pion masses as  functions of temperature and chemical potential have been investigated using the NJL model. The authors  find  that the variation of mass with respect to the chemical potential   for kaon and pion is quite different  at very small chemical potentials.
 In \cite{Ruivo}, the behavior of kaons' mass has been investigated in a flavor-asymmetric quark medium at finite temperature, within the  SU(3) NJL model.  It was shown that the  $K^{+}$ mass raises with increasing the baryonic density for different temperatures whereas $K^{-}$ mass decreases. The splitting between the  $K^{+}$ and $K^{-}$ masses decreases with temperature, as well. In \cite{Zakout}, kaon production in hot and dense hypernuclear matter has been studied via the modified quark–meson coupling model (MQMC). The authors have investigated  the kaon effective mass versus baryon density for various temperatures. Although it is expected that the effective hadronic masses tend to  vanish at large baryon density according to the standard Walecka model, the authors have observed that kaon mass for various temperatures at large baryon density decreases but it doesn't  vanish and kaon production starts smoothly at low temperatures and then they are abundantly produced when the temperature reaches the critical one. In \cite{Lavagno}, the strangeness production at finite temperature and baryon density has been investigated neglecting the contribution of the neutral kaons ($K^{0}$ and $\overline{K}^{0}$) by an effective relativistic mean-field model.   
In \cite{Tolos}, the properties of $K$ and $\overline{K}$ mesons have been studied in nuclear matter at finite temperature from a chiral unitary approach in coupled channels which incorporates the S- and P-waves of the kaon-nucleon interaction.  In \cite{Mishra}, the medium modifications of kaon and antikaon masses in hot and dense hadronic matter have been studied in a chiral SU(3) model.  The authors find  that the $K^{+}$ mass increases with the density while the $K^{-}$ mass decreases and vanishes at large densities for $T=150$ MeV. In \cite{Bhattacharyya}, the masses and decay widths of pion and kaon  in a medium of hot quark matter have been studied in the NJL model. It has been observed that their masses remain roughly unchanged up to $T=200$ MeV but increases sharply after this point. In \cite{Iazzi}, the kaons production at finite temperature and baryon density has been investigated via an effective relativistic mean-field model with the inclusion of the full octet of baryons. In \cite{Blaizot}, the $K$ and $\phi$ masses at finite temperature have  been calculated in the NJL model.

In this paper, we investigate the behaviour of the mass and decay constant of the strange particle kaon at finite temperature and baryon chemical potential using QCD sum rules. Information obtained from our numerical analysis allow us to comment on the QCD phase diagram in the $T-\mu_{B}$ plane. 

The organization of the paper is as follows: In Sec. II, we construct QCD sum rules for the mass and decay constant of kaon at finite temperature and density. Sec. III is devoted to numerical analyses of the obtained sum rules. Sec IV is reserved for concluding remarks.

\section{Temperature and baryon chemical potential dependent sum rules}

\begin{figure}[ht]
\begin{center}
\includegraphics[width=7.5cm]{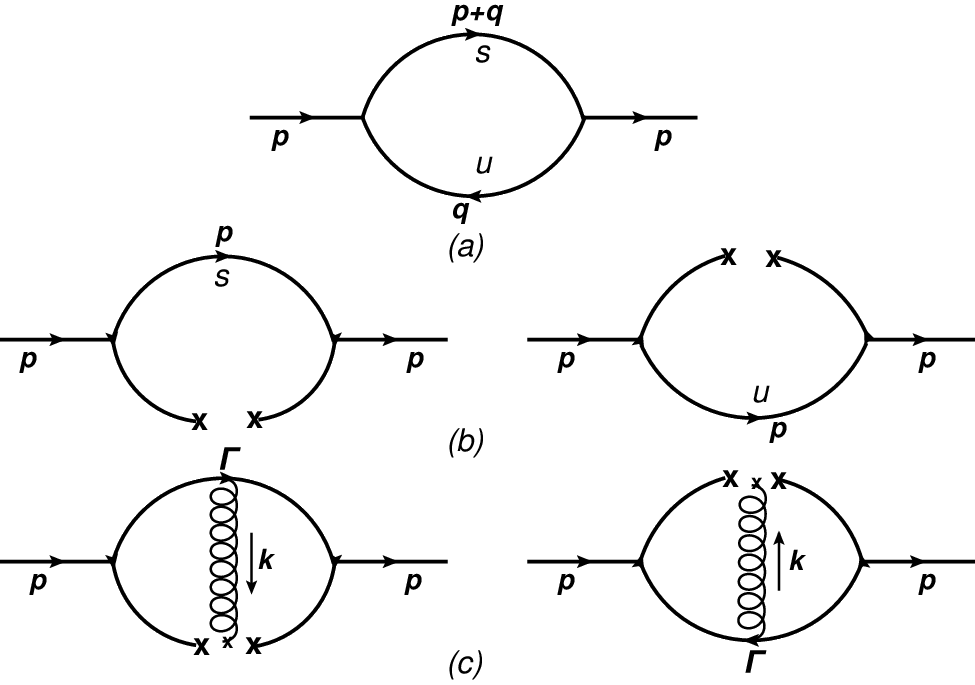}
\end{center}
\caption{(a) Bare loop diagram for perturbative contribution; (b) Diagrams corresponding to quark condensate with dimension three; (c) Mixed condensate diagrams with dimension five.}\label{fig}
\end{figure}

In this section, the dependence of the properties of kaon, a light strange pseudoscalar particle with spin-parity $ J^{P}=0^{-} $, on temperature T and baryon chemical potential $ \mu_{B} $ is examined using the QCD sum rules. To this end, the contributions of temperature and baryon chemical potential dependent perturbative and non-perturbative operators  with three and  five mass dimensions given in diagrams of Fig.  (\ref{fig}) are taken into account.

Our main goal in the present section is to construct sum rules for the mass and decay constant of kaon in the hot and dense medium by calculating an appropriate  two-point correlation function in two different momentum regions. To obtain the desired sum rules, the coefficients of the selected structures from both regions are equated to each other using some assumptions. In deep Euclidean region, $ p^{2}\ll-\Lambda^{2}_{QCD} $, the calculation is made via Wilson or operator product expansion (OPE), where the perturbative (short distance effects; part (a) in Fig.  (\ref{fig}) ) and the non-perturbative (long distance effects;  parts (b) and (c) in Fig.  (\ref{fig}) ) contributions are separated. The result obtained by this way is called QCD side which is presented in terms of QCD degrees of freedom.  It is given in the form,
\begin{eqnarray}\label{QCD}
\Pi_{\mu \nu }^{QCD}(p^{2},\mu_{B}, T)=\Pi_{\mu \nu }^{pert.}(p^{2},\mu_{B}, T)+\Pi_{\mu \nu }^{np.}(p^{2},\mu_{B}, T),\notag \\
\end{eqnarray}
where  $ \mu $ and $ \nu $ are Lorentz indices and $p=(p_{0}+\mu_{B},\overrightarrow {p}) $  is the four-momentum of the particle.
While the non-pertubative contributions are expressed in terms of the thermal and dense expected values of the quark and gluon condensates, the perturbative temperature and baryon chemical potential dependent contributions are obtained by the perturbation theory. The perturbative part of correlator is defined in terms of a dispersion integral. Hence,
\begin{equation}\label{Dispersion}
\Pi_{\mu \nu }^{QCD}(p^{2},\mu_{B}, T)=\int ds\frac{\rho_{\mu \nu }(s,\mu_{B},T)}{s-p^{2}}+\Pi_{\mu \nu }^{np.}(p^{2},\mu_{B}, T),
\end{equation}
where $ \rho_{\mu \nu }(s,\mu_{B},T) $ is the spectral density given by
\begin{equation}\label{Spectral}
\rho_{\mu \nu }(s,\mu_{B},T)=\frac{1}{\pi}Im \Pi_{\mu \nu }^{pert.}(p^{2},\mu_{B}, T) \tanh \Big(\beta\frac{\sqrt{s}+\mu_{B}}{2}\Big).
\end{equation}
Here $ \beta $ is inverse of the temperature, $ \beta=1/T$, and we work in the rest frame of the initial particle, $ \overrightarrow {p}=0 $, and $ p_{0}=\sqrt{s}$. We will calculate the perturbative spectral density and non-perturbative contributions of QCD side later.

It is possible to figure out the same correlator in terms of the hadronic parameters such as the modified mass and decay constant of the considered particle. This representation of the  correlator is so-called the physical or hadronic side. To obtain the physical side, the starting point is to write the following  two-point correlator:
\begin{equation}\label{CorrFuncPhys}
\Pi_{\mu \nu}^{Phys.}(p,\mu_{B},T)=i\int d^{4}x~e^{ip\cdot x}\langle
\Psi|\mathcal{T}\{J_{\mu}(x) {J}_{\nu}^{\dag}(0)\}|\Psi\rangle,
\end{equation}
 where $\Psi $ symbolizes the hot and dense medium and $\mathcal{T}$ stands for the time-ordered product. $J_{\mu}(x)=:\bar{s}^{a}(x)\gamma_{\mu}\gamma_{5}u^{a}(x): $ is the interpolating current for kaon. The current chosen for the kaon couples to both the pseudoscalar (PS) and axial vector (AV) kaon channels, simultaneously. To obtain the physical side of the correlator, a complete set of the corresponding particle state is placed between the interpolating currents. After performing four-integral over $ x $, we obtain
\begin{eqnarray}\label{physical}
\Pi_{\mu \nu}^{Phys.}(p,\mu_{B},T) &=& \frac{\langle
	\Psi|J_{\mu}|K_{AV}\rangle \langle K_{AV}|J_{\mu}^{\dagger}|\Psi\rangle}{m^{*2}_{AV}-p^{2}} \notag \\ 
&+&\frac{\langle
	\Psi|J_{\mu}|K_{PS}\rangle \langle K_{PS}|J_{\mu}^{\dagger}|\Psi\rangle}{m^{*2}_{PS}-p^{2}}+~\cdots,\notag \\
\end{eqnarray}
where the ellipsis indicate the contributions of the higher and continuum states in both channels and $ m^{*}=m(\mu_{B},T) $. The matrix elements given in Eq. (\ref{physical}) can be written in terms of the hadron mass and decay constant as follows
\begin{eqnarray}\label{matrix}
&&\langle \Psi|J_{\mu}|K_{AV}\rangle= m_{AV}^{*}f_{AV}^{*}\varepsilon_{\mu}^{(\lambda)} \notag \\ &&
\langle
\Psi|J_{\mu}|K_{PS}\rangle=m_{PS}^{*}f_{PS}^{*}p_{\mu},
\end{eqnarray}
where  $ f^{*}=f(\mu_{B},T) $ and $ \varepsilon_{\mu}^{(\lambda)} $ is polarization vector of the axial vector state of kaon satisfying the relation
\begin{eqnarray}\label{pol}
 \sum_{\lambda}\varepsilon_{\mu}^{(\lambda)}\varepsilon_{\nu}^{*(\lambda)}=-g_{\mu\nu}+\frac{p_{\mu}p_{\nu}}{m_{AV}^{*2}}.
\end{eqnarray}
Substituting  Eqs. (\ref{matrix}) and (\ref{pol}) in Eq. (\ref{physical}), we get the following simple expression in terms of different Lorentz structures
\begin{eqnarray}\label{physical1}
\Pi_{\mu \nu}^{Phys.}(p,\mu_{B},T) &=& \frac{m_{AV}^{*2}f_{AV}^{*2}}{m^{*2}_{AV}-p^{2}}\Big(-g_{\mu\nu}+\frac{p_{\mu}p_{\nu}}{p^{2}}\Big) \notag \\ &+&
\frac{m_{PS}^{*2}f_{PS}^{*2}}{m^{*2}_{PS}-p^{2}}p_{\mu}p_{\nu}+...~.
\end{eqnarray}
In this study, we only consider the contribution of the pseudoscalar state. Therefore, we need to eliminate the pollution from the axial vector state. This undesirable contribution can be removed by multiplying the physical side by $ p_{\mu}p_{\nu}/p^{2} $. Besides, in order to isolate and enhance the ground state contribution, contributions of the higher states and continuum should be sufficiently suppressed. To do this, the Borel transformation given below is used:
\begin{eqnarray}\label{physical1}
\hat{\textbf{B}}[f(Q^{2})]=\lim_{\stackrel{Q^{2},n\rightarrow\infty}{\frac{Q^{2}}{n}=M^{2}}}\frac{(-1)^{n-1}}{(n-1)!}(Q^{2})^{n}\Big(\frac{d}{dQ^{2}}\Big)^{n-1}f(Q^{2}),\notag \\
\end{eqnarray}
where $ Q^{2}=-p^{2} $ and a new variable called the Borel mass parameter, $ M^{2} $, is introduced. Consequently, we get
\begin{eqnarray}\label{Borel}
\hat{\textbf{B}} \Pi_{\mu\nu}^{Phys.}(\mu_{B},T,M^{2})&=&m_{PS}^{*4}f_{PS}^{*2}exp[-\frac{m_{PS}^{*2}}{M^{2}}] \textbf{I}+..., \notag \\
\end{eqnarray} 
where $ \textbf{I} $ denotes the unit matrix.

The second stage of the calculations is to define the QCD side of the correlator  based on the local OPE of the quark and gluon fields. This was introduced by K.G. Wilson in Ref. \cite{Wilson} and its utilization to the correlator gives:
\begin{eqnarray}\label{OPE}
\Pi_{\mu \nu}^{QCD}(p,\mu_{B},T)&=&i\int d^{4}x~e^{ip\cdot x}\langle
\Psi|\mathcal{T}\{J_{\mu}(x) {J}_{\nu}^{\dag}(0)\}|\Psi\rangle \notag \\
&=& C_{0}\textbf{I}+\sum_{d}C_{d}(x)\langle\Psi|\hat{O}_{d}(0)|\Psi\rangle,
\end{eqnarray} 
where $C_{d}(x)$ are Wilson coefficients and  $\hat{O}_{d}$ denote a set of local operators increasing according to their dimension $  d$ in units of mass. The first term on the right-hand side of the Eq. (\ref{OPE}) is expressed by the lowest-dimensional operator $  d=0$, which corresponds to the unit matrix, and represents the perturbative contribution. The remeaning operators in the series represent non-perturbative contributions. Note that there is no colorless operator in dimension $d = 1, 2 $ that contributes to OPE in QCD. Therefore, the lowest non-perturbative operator is the quark condensate with $ d=3 $. As we previously said, in our calculations, the correlator is investigated by considering the contributions of operators up to five dimensions, which is the mixed condensate, $\langle\bar{q}g_{s}\sigma Gq\rangle=m_{0}^{2}\langle\bar{q}q\rangle $.

The computation of the QCD side is started with the calculation of the perturbative contribution to obtain the corresponding spectral density. The  amplitude for the bare loop is written as 
 %.The main contribution to the QCD part of correlator come from bare loop diagram.
%
\begin{equation}\label{CorrFunc}
\Pi_{\mu \nu }^{pert.}(p,\mu_{B}, T)=i N_{c}\int \frac{d^{4}q}{(2 \pi)^{4}} Tr \Big[S(q)\gamma_{\nu}\gamma_{5} S(p+q)\gamma_{\mu}\gamma_{5} \Big],
\end{equation}
where $ N_{c} = 3 $ is the color factor and the dependence on temperature is expressed explicitly by the Fermi-Dirac distribution function in the thermal quark propagator \cite{tez} given by
\begin{equation}\label{propagator}
S(q)=i(\!\not\!{q}+m)\Big[ \frac{1}{q^{2}-m^{2}+i\varepsilon} +2\pi i n_{f}(| q_{0}|) \delta(q^{2}-m^{2}) \Big],
\end{equation}
 with the Fermi-Dirac distribution function defined as $n_{f}(x)=(e^{\beta x}+1)^{-1}$. The dependence on chemical potential is defined within the external four-momentum, $p=(p_{0}+\mu_{B},\overrightarrow {p})$. Now, we insert the expression of the propagator given in Eq. (\ref{propagator}) into the correlation function in Eq. (\ref{CorrFunc}) and we obtain
%$ {\textbf p}^{2}=(p_{0}+\mu)^{2} -\overrightarrow {p}^{2}$.
% 
\begin{eqnarray}\label{CorrFunc1}
&&\Pi_{\mu \nu }^{pert.}(p,\mu_{B}, T)=-i N_{c}\int \frac{d^{4}q}{(2 \pi)^{4}}\nonumber \\
&\times& Tr \Big((\!\not\!{q}+m_{u})\gamma_{\nu}\gamma_{5} (\!\not\!{p}+\!\not\!{q}+m_{s})\gamma_{\mu}\gamma_{5} \Big)\nonumber \\
&\times&\Big[\frac{1}{q^{2}-m_{u}^{2}+i\varepsilon} \frac{1}{(p+q)^{2}-m_{s}^{2}+i\varepsilon}\nonumber \\
&+&\frac{1}{q^{2}-m_{u}^{2}+i\varepsilon} 2\pi i n_{f}(| p_{0}+\mu_{B}+q_{0} |) \delta((p+q)^{2}-m_{s}^{2})\nonumber \\
&+&\frac{1}{(p+q)^{2}-m_{s}^{2}+i\varepsilon} 2\pi i n_{f}(| q_{0} |) \delta(q^{2}-m_{u}^{2})\nonumber \\
&-&4\pi^{2} n_{f}(|q_{0}|) n_{f}(|p_{0}+\mu_{B}+q_{0}|)\delta(q^{2}-m_{u}^{2})
\nonumber \\
&\times&
\delta((p+q)^{2}-m_{s}^{2})\Big].\nonumber \\
\end{eqnarray}
To express the calculations in details, we can write Eq. (\ref{CorrFunc1}) in four parts. The calculations of the first part is clearly expressed as follows: 
\begin{eqnarray}\label{QCDpart1}
&&\Pi_{\mu\nu,1 }^{pert.}(p,\mu_{B}, T)=-i N_{c}\int \frac{d^{4}q}{(2 \pi)^{4}}\notag \\
&\times&
 Tr \Big((\!\not\!{q}+m_{u})\gamma_{\nu}\gamma_{5} (\!\not\!{p}+\!\not\!{q}+m_{s})\gamma_{\mu}\gamma_{5} \Big)
\notag \\
&\times&\Big[\frac{1}{q^{2}-m_{u}^{2}+i\varepsilon} \frac{1}{(p+q)^{2}-m_{s}^{2}+i\varepsilon}\Big].
\end{eqnarray}
After performing the trace, we get
\begin{eqnarray}\label{}
&&\Pi_{\mu\nu,1 }^{pert.}(p,\mu_{B}, T)=- 4iN_{c}\int \frac{d^{3}\overrightarrow{q}}{(2 \pi)^{4}}\int dq_{0}\notag \\
&\times&
 \Big(p_{\nu}q_{\mu}+p_{\mu}q_{\nu}+2q_{\mu}q_{\nu}-(p\cdot q+q^2+m_{u}m_{s})g_{\mu\nu} \Big)
\notag \\
&\times&\Big[\frac{1}{q^{2}-m_{u}^{2}+i\varepsilon} \frac{1}{(p+q)^{2}-m_{s}^{2}+i\varepsilon}\Big].
\end{eqnarray}
Performing the Feynman integrals with the help of the Cutkosky rules, which allows us to change the quark propagator with the Dirac delta function representing the related quark to be real,
\begin{eqnarray}\label{fermi}
\frac{1}{q^{2}-m^{2}+i\varepsilon}=(-2i \pi)\delta(q^{2}-m^{2}),
\end{eqnarray}
 we get the first part of the perturbative term as
 \begin{eqnarray}\label{QCDpart1}
 &&\Pi_{1}^{pert.}=\frac{-3i(m_{u}+m_{s})^{2}}{8\pi (\sqrt{s}+\mu_{B})^{4}}\varDelta^{2}\sqrt{1-\frac{4m_{u}m_{s}}{\varDelta}},
 \end{eqnarray}
where $ \varDelta=(\sqrt{s}+\mu_{B})^{2}-(m_{u}-m_{s})^{2} $. It should be remembered that the above equation was obtained by multiplying the result with $ p_{\mu}p_{\nu}/p^{2} $ to get only the pseudoscalar contribution like the physical side. The other three parts are calculated similarly. As a result, we get
\begin{equation}\label{QCDpart1total}
\Pi^{pert.}=\Pi_{1}^{pert.}\Big[1-n_{f}(|\omega_{1}|)-n_{f}(|\omega_{2}|)+2n_{f}(|\omega_{1}|)n_{f}(|\omega_{2}|)\Big],
\end{equation}
where $ \omega_{1}=\frac{m_{u}^{2}-m_{s}^{2}+(\sqrt{s}+\mu_{B})^{2}}{2(\sqrt{s}+\mu_{B})} $ and $ \omega_{2}=(\sqrt{s}+\mu_{B})-\omega_{1} $. To get the final form of the spectral density, we need to take the imaginary part of Eq. (\ref{QCDpart1total}) and multiply it with tangent hyperbolic function expressed in Eq.(\ref{Spectral}). After all these standard calculations, we obtain the temperature and chemical potential dependent spectral density as
\begin{eqnarray}\label{SD}
\rho(s,\mu_{B},T)&=&\frac{-3(m_{u}+m_{s})^{2}}{8\pi^{2} (\sqrt{s}+\mu_{B})^{4}}\varDelta^{2}\sqrt{1-\frac{4m_{u}m_{s}}{\varDelta}}\notag \\
&\times&(1-n_{f}(|\omega_{1}|)-n_{f}(|\omega_{2}|)).
\end{eqnarray}

To proceed, we also need to calculate the contribution of the non-perturbative terms on the QCD side. For this aim, we consider the quark and mixed condensate diagrams presented in (b) and (c) parts of Fig.  (\ref{fig}) and implement the medium effects through the condensates.
% Because of the insignificant small contribution of gluon condensate,  $ \langle G^{2} \rangle $, we may ignore these terms and we do not present their explicit expressions.
 In the momentum space, the correlator for the contributions of the quark and mixed condensates can be written as
\begin{eqnarray}\label{}
&&\Pi_{\mu\nu}^{np.}(p,\mu_{B}, T)=\langle\Psi|\bar{u}_{\alpha}^{i}(0)\Big(\gamma_{\nu}\gamma_{5}S(p)\gamma_{\mu}\gamma_{5}\Big)_{\alpha\beta}u_{\beta}^{j}(x)|\Psi\rangle,
\notag \\
\end{eqnarray}
and
\begin{eqnarray}\label{}
&&\Pi_{\mu\nu}^{np.}(p,\mu_{B}, T)=\notag \\&&  \langle\Psi|\bar{u}_{\alpha}^{i}(0)\Big(\gamma_{\nu}\gamma_{5}S(p)\Gamma S(p+k)\gamma_{\mu}\gamma_{5}\Big)_{\alpha\beta}u_{\beta}^{j}(x)|\Psi\rangle,\nonumber \\
\end{eqnarray}
respectively, where $ \Gamma=i\frac{g}{2}x_{\lambda}\gamma_{\tau}G_{\lambda\tau} $ with $x_{\lambda}=- i\frac{\partial}{\partial k_{\lambda}} $. To continue the operations, we need to expand the quark field defined at the $ x $ point to the Taylor series around $ x=0 $, i.e.

%
%\begin{equation}\label{}
%u_{\alpha}(x)=u_{\alpha}(0)+x_{\alpha}\nabla_{\alpha}u_{\alpha}(x)|_{x=0}+\frac{1}{2}x_{\alpha}x_{\beta}\nabla_{\alpha}%\nabla_{\beta}u_{\alpha}(x)|_{x=0}+...
%\end{equation}
%
%
\begin{equation}\label{}
	u_{\alpha}(x)=u_{\alpha}(0)+x_{\beta}\nabla_{\beta}u_{\alpha}(0)+\frac{1}{2}x_{\beta}x_{\beta'}\nabla_{\beta}\nabla_{\beta'}u_{\alpha}(0)+~\cdots.
\end{equation}
After standard calculations, for the non-perturbative part in Borel scheme we obtain
\begin{eqnarray}\label{}
&&\hat{B} \Pi^{np.}(\mu_{B},T,M^{2})=\notag \\ && \Big (m_{s}-m_{u}+\frac{m_{u}m_{s}^{2}}{2M^{2}}+\frac{m_{u}^{2}m_{s}^{3}}{2M^{4}}\Big)e^{\frac{-m_{s}^{2}}{M^{2}}} \langle\varPsi|\bar{u}u|\varPsi\rangle\notag \\
&+&\Big(m_{u}-m_{s}+\frac{m_{s}m_{u}^{2}}{2M^{2}}+\frac{m_{s}^{2}m_{u}^{3}}{2M^{4}}\Big)e^{\frac{-m_{u}^{2}}{M^{2}}} \langle\varPsi|\bar{s}s|\varPsi\rangle
\notag \\
&-&\frac{m_{s}^{3}m_{0}^{2}}{4M^{4}}\langle\varPsi|\bar{u}u|\varPsi\rangle e^{\frac{-m_{s}^{2}}{M^{2}}}-\frac{m_{u}^{3}m_{0}^{2}}{4M^{4}}\langle\varPsi|\bar{s}s|\varPsi\rangle e^{\frac{-m_{u}^{2}}{M^{2}}}.\nonumber \\
\end{eqnarray}
Finally, to derive the desired sum rules for the mass and decay constant, both the QCD and physical sides of the correlator are matched. Hence,
%
%\begin{eqnarray}\label{SRDC}
%f_{PS}^{*2}m_{PS}^{*4}e^{-m_{PS}^{*2}/M^2}=\varUpsilon(s_{0},\mu_{B},T,M^{2}),
%\end{eqnarray}
%
%
\begin{eqnarray}\label{SRDC}
f_{PS}^{*2}=\frac{\varUpsilon(s_{0},\mu_{B},T,M^{2})}{m_{PS}^{*4}}e^{m_{PS}^{*2}/M^2},
\end{eqnarray}
and
\begin{eqnarray}\label{SRM}
m_{PS}^{*2}=\frac{\frac{d\varUpsilon(s_{0},\mu_{B},T,M^{2})}{d(-1/M^2)}}{\varUpsilon(s_{0},\mu_{B},T,M^{2})},
\end{eqnarray}
where
\begin{eqnarray}\label{}
\varUpsilon(s_{0},\mu_{B},T,M^{2})&=&\int_{(m_{u}+m_{s})^2}^{s_{0}(\mu_{B},T)}ds\rho(s,\mu_{B},T)e^{-s/M^2}\notag \\
&+&\hat{B} \Pi^{np.}(\mu_{B},T,M^{2}),
\end{eqnarray}
and $ s_{0}=s_{0}(0,0) $. The baryon chemical potential and temperature dependent continuum threshold as well as quark condensate are given as  \cite{Ayala1}
\begin{eqnarray}\label{Conthreshold}
\frac{s_{0}(\mu_{B},T)}{s_{0}}\backsimeq \frac{\langle\varPsi| \bar{q}q|\varPsi \rangle}{\langle 0| \bar{q}q|0 \rangle}-\frac{T^{2}/3-\mu_{B}^{2}/\pi^2}{2f_{\pi}^2(0,0)},
\end{eqnarray}
and
\begin{eqnarray}\label{quarkCond}
\langle\varPsi| \bar{q}q|\varPsi \rangle&=&\langle0| \bar{q}q|0 \rangle-\frac{24T}{\pi^{2}}\sum^{\infty}_{l=1}\frac{(-1)^{l}}{l}\cosh\notag \\ 
&\times& \Big(\frac{\mu_{B}l}{T}\Big)\sum^{3}_{i=1}\frac{r_{i}m_{i}^{2}}{\mid b_{i}\mid^{3}}K_{1}\Big(\frac{l\mid m_{i}\mid}{T}\Big){\tiny }.
\end{eqnarray}
Note that Eq. (\ref{quarkCond}) is valid only in the $ 0\leq \mu_{B} < 0.50 $~ GeV for the baryon chemical potential. Here $ f_{\pi}(0,0)=92.21\pm0.14~MeV $ \cite{Nakamura},  $ b_{1,2,3}=1 $ and $ K_1(x) $ is a Bessel function. The values for parameters $ m_{i} $ and $ r_{i} $ are given in Table (\ref{tab:1}).
\begin{table}[h]
	\centering
	\begin{tabular}{|c||c|c|c|}\hline \hline
		$i$ & $1$ & $2$ & $3$   \\ \hline
	$ m_{i}~(GeV) $ & $-0.490$ & $0.495$ & $-0.879$ \\
		$ r_{i} $ &$ -0.112 $&$ 0.352 $&$ 0.259 $ \\ \hline \hline
	\end{tabular}
	\vspace{0.8cm}
	\caption{Values of parameters given in the quark condensate \cite{Ayala1}.}\label{tab:1}
\end{table}

%The sum rules of the decay constant can be seen easily in Eq. (\ref{SRDC}). But, to obtain the sum rules for the mass, %the derivative of Eq.(\ref{SRDC}) according to $ (-1/M^{2}) $ is divided by itself and the result is

%%%%%%%%%%%%%%%%%%%%%%%%%%%%%%%%%%%%%%%%%%%%%%%%%%%%%%%%%%%%%%%%%%%%%%%%%%%%%%%%%%%%%%%%%%%%%%%%%%%
\section{Numerical results}

\begin{figure}[ht]
	\begin{center}
		\includegraphics[width=7.5cm]{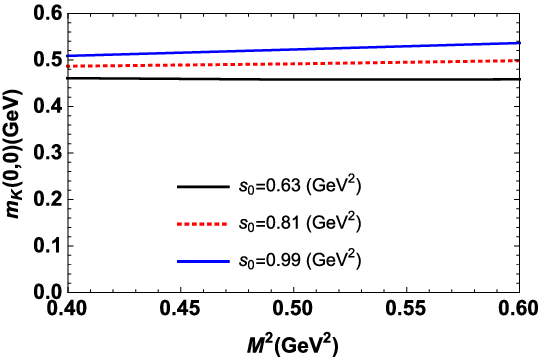}
	\end{center}
	\caption{The mass of kaon in vacuum with respect to Borel parameter at different continuum thresholds.} \label{fig2}
\end{figure}

\begin{figure}[ht] 
	\begin{center}
		\includegraphics[width=7.5cm]{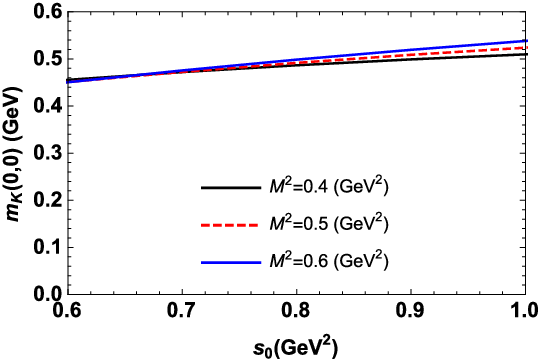}
	\end{center}
	\caption{For different Borel paremeters, the mass of kaon with respect to continuun threshold in vacuum.} \label{fig3}
\end{figure}

After obtaining the sum rules for the  mass and decay constant in a hot and dense medium, we present the numerical analysis in this section to obtain the vacuum values and dependence on temperature/baryon chemical density of physical quantities. From the acquired sum rules, we see that the quark masses, quark and mixed condensates are the main input parameters. The quark masses are taken to be $ m_{u}=(2.3_{-0.5}^{+0.7}) ~MeV $ and $  m_{s}=(93_{-5}^{+11} )~MeV$ \cite{PDG}. The values of quark condensates in vacuum are $ \langle 0| \bar{q}q|0 \rangle(1GeV)=(-0.24\pm0.01)^3~ GeV^3 $ and $ \langle 0| \bar{s}s|0 \rangle=0.8\langle 0| \bar{q}q|0 \rangle $. The mixed condensates are $\langle 0| \bar{q}g_{s}\sigma Gq|0 \rangle=m_{0}^{2} \langle 0| \bar{q}q|0 \rangle $ and $\langle 0| \bar{s}g_{s}\sigma Gs|0 \rangle=m_{0}^{2} \langle 0| \bar{s}s|0 \rangle $ with $ m_{0}^{2}=(0.8\pm0.1)~GeV^{2} $ \cite{Belyaev}. In addition to input parameters, the working windows of two extra auxiliary parameters entering the calculations are needed: the continuum thresholds $ s_0  $ and the  Borel parameter $ M^2 $.  We look for some windows of these helping parameters such that  the results of physical observables depend relatively weakly on these parameters. To this end, some standard criteria of the method  such as the pole dominance over the higher states and continuum and convergence of the OPE are applied. The continuum threshold is not completely arbitrary and it is related to the energy of the first exited state of the  particle under consideration. Thus, for  the intervals of the Borel parameter and continuum threshold, we obtain, $ M^{2}\in[0.4-0.6]~GeV^{2} $ and $ s_{0}\in [0.63-0.99]~ GeV^{2}$. 
%Our numerical analysis satisfy the interval of Borel parameter are $ M^{2}\in[0.4-0.6]~GeV^{2} $ and continuum threshold are $ s_{0}\in [0.63-0.99]~ GeV^{2}$. 
The graphs given in Figs. (\ref{fig2}) and  (\ref{fig3}) are plotted to check the reliability of the selected working regions. These figures depict that, in the selected working regions, the dependence of the mass of kaon on auxiliary parameters is weak: The residual dependencies appear as parts of the uncertainties in the results. Using the input parameters as well as the  obtained intervals of Borel parameter and continuum threshold, the vacuum values of mass and decay constant for kaon were obtained as

\begin{eqnarray}\label{vacuum}
m_{K}(0,0)=491.7^{+44.5}_{-31.0}~MeV \notag \\ 
f_{K}(0,0)=157.2^{+22.0}_{-25.2}~MeV .
\end{eqnarray}
These values  are nicely consistent with both the existing experimental data  as well as other phenomenological models predictions \cite{PDG,Er14,Lattice,Lattice1,Lattice2,Lattice3}.

\begin{figure}[ht] 
	\begin{center}
		\includegraphics[width=7.5cm]{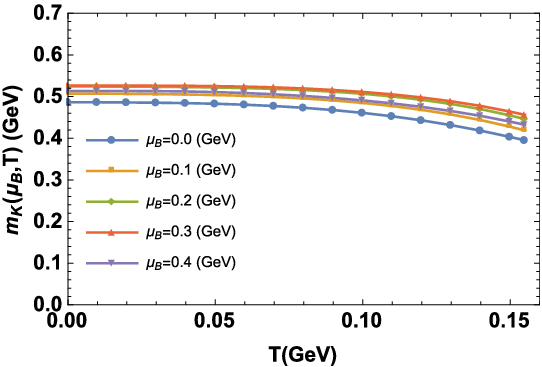}
	\end{center}
	\caption{Temperature-dependent mass for different baryon chemical potential.} \label{fig4}
\end{figure}

\begin{figure}[ht] 
	\begin{center}
		\includegraphics[width=7.5cm]{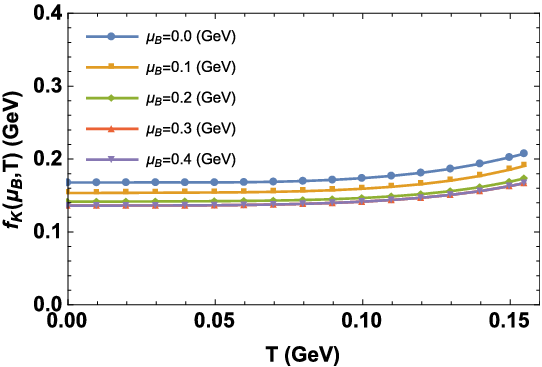}
	\end{center}
	\caption{Temperature-dependent decay constant for different baryon chemical potential.}\label{fig5}
\end{figure}

\begin{figure}[ht]
	\begin{center}
		\includegraphics[width=7.5cm]{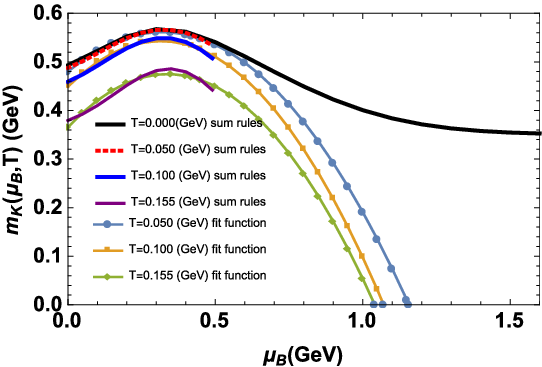}
	\end{center}
	\caption{Mass-baryon chemical potential graph for different temperatures.}\label{fig6}
\end{figure}

\begin{figure}[ht]
	\begin{center}
		\includegraphics[width=7.5cm]{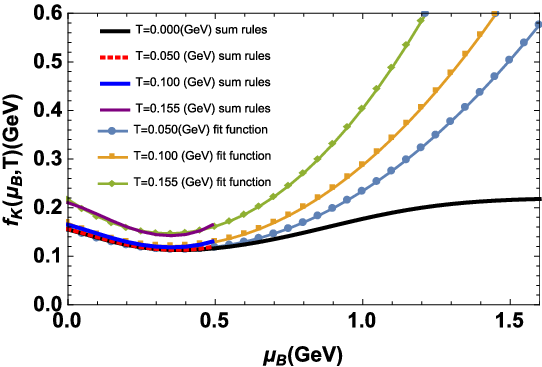}
	\end{center}
	\caption{Decay constant-baryon chemical potential graph for different temperatures.
		}\label{fig7}
\end{figure}

The main goal of this section is to investigate the dependence of the physical observables on baryon chemical potential and temperature. For this purpose, we first plot the temperature-dependent graphs of the mass and the decay constant for different fixed values of the chemical potential  in Figs. (\ref{fig4}) and  (\ref{fig5}).
Looking at the mass-temperature graph, the mass does not change in the temperature range $[0-50] $  MeV at fixed  chemical potential values, but after this point, it is seen that the mass decreases as the temperature increases. The amount of reduction in the value of mass at critical temperature  is about $ (14-19)\% $. 
%At the same time, while the mass values are the same between $ \mu_{B}=[0-100]~MeV $, with the increase of the baryon chemical potential from this value, the mass decreases for each $ \mu_{B}$, even if it is small. 
For the decay constant-temperature graph, the values are stable in the range $ [0-70] $  MeV of the temperature, after which an  increase about  $ (18-20)\% $ is seen at the critical temperature.  Figs. (\ref{fig6}) and (\ref{fig7}) are drawn to show the dependence of the physical quantities under study on baryon chemical potential for certain temperature values. As it is seen from these figures,  except for zero temperature, we get reliable results only for the $ 0\leq \mu_{B} < 0.50 $~ GeV at non-zero temperatures because of the used quark condensate expression, which is limited to the lower values of the baryon chemical potential. To  extrapolate the results to higher values of $ \mu_{B} $, we shall use some  fit functions.
 Various fit functions  for this purpose can be suggested: The most suitable function among them is obtained as:
\begin{eqnarray}\label{fitmass}
m_{K}[f_{K}](\mu_{B})&=&\theta \mu_{B}^{2} +\delta \mu_{B}+\xi,
\end{eqnarray}  
where $ \theta, \delta $ and $ \xi $ are fitting parameters  collected in Table (\ref{tab:2}) for both the mass and decay constant at fixed temperatures. As we noted above there are several fit functions to extrapolate the results to the higher chemical potentials, which impose some uncertainties to the results. We take into account these uncertainties by adding the related errors to the parameters in Table (\ref{tab:2}).
\begin{table}[h]
	\centering
	\begin{tabular}{|c||c|c|c|}\hline \hline
		$ m_{K} $  \\ \hline 
		$ $ & $\theta~(1/GeV)$ & $\delta$ & $\xi~(GeV)$   \\ \hline
		$ T=0.050~(GeV) $ & $-0.799^{+0.140}_{-0.252}$ & $0.510^{-0.070}_{+0.108}$ & $0.480^{+0.012}_{-0.026}$ \\
		$ T=0.100~(GeV) $ & $-0.948^{+0.140}_{-0.188}$ & $0.596^{-0.066}_{+0.097}$ & $0.449^{+0.012}_{-0.019}$ \\
		$ T=0.155~(GeV) $ &$ -0.961^{+0.148}_{-0.244} $&$ 0.649^{-0.065}_{+0.083} $&$ 0.365^{+0.021}_{-0.020} $ \\ \hline \hline
		$ f_{K} $  \\ \hline 
		$ $ & $\theta~(1/GeV)$ & $\delta$ & $\xi~(GeV)$   \\ \hline
		$ T=0.050~(GeV) $ & $0.309^{+0.009}_{-0.027}$ & $-0.233^{+0.006}_{-0.020}$ & $0.158^{+0.004}_{-0.014}$ \\
		$ T=0.100~(GeV) $ & $0.393^{+0.011}_{-0.035}$ & $-0.274^{+0.008}_{-0.024}$ & $0.168^{+0.005}_{-0.015}$ \\
		$ T=0.155~(GeV) $ &$0.596^{+0.017}_{-0.053}$&$ -0.409^{+0.012}_{-0.036} $&$0.217^{+0.006}_{-0.019} $ \\ \hline \hline		
	\end{tabular}
	\vspace{0.8cm}
	\caption{Coefficients of the  fit function for the mass and decay constant in Eq. (\ref{fitmass}) for different temperatures. The  superscripts and  subscripts in the presented values provide  bands for the mass and decay constant that become wider at higher values of the baryon chemical potential.}\label{tab:2}
\end{table}
 Figs. (\ref{fig6}) and (\ref{fig7})  show  reasonable agreements of the fitting results with QCD sum rules predictions at small chemical potentials.  These figures show that the mass/decay constant increases/decreases up to roughly  $ \mu_{B}=0.4 $, GeV at constant temperatures, however, after this point, the mass/decay constant starts to decrease/increase. The mass  apparently vanishes in the interval $\mu_{B}=(1.03-1.15)$~ GeV for non-zero temperatures. It can be considered as a sign for the deconfinement of the hadronic state and phase transition  to the QGP. Both the mass and decay constant reach to a constant value at zero temperature and higher baryon chemical potential. 
  The point of apparent vanishing of mass at non-zero temperatures  moves to lower baryon chemical potentials by increasing the temperature. This behavior (except for zero temperature) is  consistent  with the  QCD phase diagram in  $ T-\mu_{B}$ plane  presented in Fig. (\ref{fig1}).

%

%The solid line describes the sum rules predictions, other line denotes the prediction of fit function obtained using the central values of the input parameters
It would be instructive to give the mass and decay constant of kaon as functions of both  the temperature and chemical potential simultaneously. They read
\begin{eqnarray}\label{fitmf}
m_{K}[f_{K}](\mu_{B},T)&=&a \mu_{B}^{2} T^{2}+b \mu_{B}^{2} T+c \mu_{B} T^{2}+d \mu_{B}^{2} \nonumber\\
&+&e T^{2}+f \mu_{B} T+g T+h \mu_{B} +k.\nonumber\\
\end{eqnarray}  
Note that this parameterizations give the consistent  vacuum values given in Eq. (\ref{vacuum}) for the mass and decay constant, however,   the coefficients  are obtained using graphs for  $ T\neq0 $  in Figs. (\ref{fig6}) and (\ref{fig7}). The values of the coefficients in Eq. (\ref{fitmf}) are given in Table (\ref{tab:3}). These fit functions may be used to investigate other behaviors  of the kaon in a medium with extreme conditions.
\begin{table}[h]
	\renewcommand{\arraystretch}{1.5}
	\addtolength{\arraycolsep}{3pt}
	$$
	\begin{array}{|c|c|c|c|c|c|c|c|c|c|}
	\hline \hline
	& m_{K}(\mu_{B},T) & f_{K}(\mu_{B},T)  \\
	\hline
	\mbox{a $(GeV^{-3})$} &21.749^{+1.957}_{-3.914}&18.158^{+0.544}_{-1.634} \\
	\hline
	\mbox{b $(GeV^{-2})$} &-6.077^{+0.546}_{-1.093}&-0.970^{+0.029}_{-0.087} \\
	\hline
	\mbox{c $(GeV^{-2})$} &-7.262^{+0.653}_{-1.307}&-12.476^{+0.374}_{-1.122} \\
	\hline
	\mbox{d $(GeV^{-1})$} &-0.553^{+0.049}_{-0.099}&0.310^{+0.009}_{-0.027} \\
	\hline
	\mbox{e $(GeV^{-1})$} &-6.188^{+0.556}_{-1.113}&4.329^{+0.129}_{-0.389} \\
	\hline
	\mbox{f $(GeV^{-1})$} &2.930^{+0.263}_{-0.527}&0.839^{+0.025}_{-0.075} \\
	\hline
	\mbox{g} &0.135^{+0.012}_{-0.024}&-0.304^{+0.009}_{-0.027} \\
	\hline
	\mbox{h} &0.378^{+0.034}_{-0.068}&-0.238^{+0.007}_{-0.021} \\
	\hline
	\mbox{k $(GeV)$} &0.493^{+0.044}_{-0.088}&0.158^{+0.004}_{-0.014} \\
	\hline \hline
	\end{array}
	$$
	\caption{ The values of coefficients $a, b, c, d, e, f, g, h$ and $k$ in the expression of fit functions for  $m_{K}[f_{K}](\mu_{B},T)$.} \label{fitfunction1}
	\renewcommand{\arraystretch}{1}
	\addtolength{\arraycolsep}{-1.0pt} \label{tab:3}
\end{table}
As is seen from Figs. (\ref{fig4}),  (\ref{fig5}),  (\ref{fig6}) and  (\ref{fig7}) the behavior of the mass/decay constant of kaon in terms of temperature  is monotonic,  while it is non-monotonic with respect to the chemical potential. This neither refers to any thermodynamic instability nor presence of heavier strange particles in the system as they  don't couple to the considered current. The non-monotonic behavior of the physical quantities with respect to the chemical potential just belongs to the variations of values of the quark condensations with respect to the density. The vanishing  of  the mass at higher chemical potentials and  non-zero temperatures can be considered as a sign for phase transition to possible QGP.
%%%%%%%%%%%%%%%%%%%%%%%%%%%%%%%%%%%%%%%%%%%%%%%%%%%%%%%%%%%%%%%%%%%%%%%%%%%%%%%%%%%%%%%%%%%%%%%%%%
\section{Summary and Conclusions}
In this article, the  mass and decay constant of  the strange pseudoscalar particle kaon were investigated  in the hot and dense medium using the temperature and baryon chemical potential dependent QCD sum rules analyses.  Such investigations can provide inputs to future related experiments aiming to study the hadronic properties at hot and dense medium. These investigations can also provide hints to possible hadronic phase transitions at different temperature and baryon chemical potentials as well as get knowledge on the QCD phase diagram at $ T- \mu_{B} $ plane.

To calculate the mass and decay constant in terms of $ T $ and $ \mu_{B} $, first, we  obtained the perturbative spectral density in terms of these parameters. Then we included  the temperature  and baryon chemical potential dependent non-perturbative contributions up to operators of mass dimension five. We matched both the QCD and hadronic sides to calculate the temperature and chemical potential dependent mass and decay constant in terms of QCD degrees of freedom and other auxiliary parameters. As the chosen current couples simultaneously to both the pseudo scalar and axial vector states, we removed the contributions of axial vectors to find the desired behaviors of the parameters of the  pseudoscalar kaon. To this end, we  multiplied both sides of the sum rules  by $ p_{\mu}p_{\nu}/p^{2} $. After fixing the working intervals of the auxiliary parameters, we numerically analyzed the obtained sum rules.

The vacuum values of mass and decay constant of kaon are obtained as $ 491.7^{+44.5}_{-31.0}$ MeV and $ 157.2^{+22.0}_{-25.2} $ MeV, respectively. These results are in nice consistency with the existing experimental data as well as the Lattice and previous QCD sum rules results \cite{PDG,Er14,Lattice,Lattice1,Lattice2,Lattice3}.
% Our results on the vacuum mass and decay constant of kaon are in nice consistency with the existing experimental data as well as the Lattice and previous QCD sum rules results.
When we look at the dependence of the physical quantities on the temperature,  we see that, at fixed values of baryon chemical potentials, the mass and decay constant are not affected by the increasing in the values of temperature up to roughly  $T=50$ MeV and  $T=70$ MeV  respectively, however  after these points, the mass starts to diminish and the decay constant starts to rise up to a critical temperature  $T=155$ MeV  (obtained from  Lattice QCD calculations) considerably. In  the critical temperature, a $  (14-19)\% $ decrease in mass is observed, while the decay constant increases by $ (18-20)\% $. Comparing the    obtained result  with that of  \cite{Lavagno}, we see that the behavior of the mass with respect to temperature at a fixed baryon density is in agreement with the prediction of \cite{Lavagno}. The numerical results also demonstrate that the mass/decay constant of the particle rises/falls considerably by increasing the baryon chemical potential at  zero and finite temperatures up to approximately $\mu_{B}=0.4$ GeV. After this point, the mass/decay constant  starts to fall/increase by increasing the baryon chemical potential. The behavior of the kaon's mass in terms of the baryon chemical potential in Fig. (\ref{fig6}) is in good agreement with those presented in \cite{Ruivo,Mishra} for $\mu_{B}<0.4$  GeV   at $T\neq0$.  In principle, there may be sizable  splitting between the $K^{+}$ and $K^{-}$ masses at dense medium. It can be studied by separating the contributions of the odd and even dimensional operators in nuclear matter  \cite{Zschocke,Suzuki,Song}.   Although such splitting is moderately reduced by incensing in the temperature as shown in \cite{Ruivo}, it would be interesting to formulate and study possible splitting at finite temperature and chemical potential in pseudoscalar kaon channel. At zero temperature, the kaon mass-chemical potential graphic  is in good agreement with \cite{Mishra}, as well. It is seen that the mass apparently vanishes at $\mu_{B}=(1.03-1.15)$ GeV for finite temperatures: The point of apparent vanishing moves to lower baryon chemical potentials by increasing the temperature. At zero temperature, both the mass and decay constant reach to roughly  fixed values at higher baryon chemical potentials. The behavior of the mass with respect to the temperature and baryon chemical potential and its melting at higher baryon chemical potentials, which can be considered as a sign to transition to QGP, is  consistent with the QCD phase diagram in the $T-\mu_{B}$ plane.

We provided the functions of the mass and decay constant in terms of the $ T $ and $ \mu_{B} $. These results may be used in investigation of  other interaction and decay properties of pseudoscalar kaon at extreme conditions. Our results may help experimental groups to analyze their data in future heavy ion collision or in-medium experiments.

\section*{ACKNOWLEDGEMENTS}
K. Azizi is thankful to Iran Science Elites Federation (Saramadan)
for the partial  financial support provided under the grant number ISEF/M/401385.

%%%%%%%%%%%%%%%%%%%%%%%%%%%%%%%%%%%%%%%%%%%%%%%%%%%%%%%%%%%%%%%%%%%%%%%%%%%%%%%%%%%%%%%%%%%%%%%%%%%
%\section{Appendix}
%The Borel transformation given in Eq.(\ref{Borel}) leads to an exponential suppression of the excited states and %removes the polynomial subtraction terms. Borel transforms of functions that are polynomials are equal to zero, and %Borel transforms for some useful functions as follows
%
%\begin{eqnarray}\label{}
%&&\hat{\textbf{B}}[(Q^{2})]^{k}=0~~~~~~~~k\geq 0
%\notag \\ 
%&&\hat{\textbf{B}}\Big(\frac{1}{Q^{2}})^{k}=\frac{1}{(k-1)\textexclamdown}(\frac{1}{M^{2}})^{k}
% \notag \\ 
%&&\hat{\textbf{B}}[\frac{1}{s+Q^{2}}]^{k}=\frac{1}{(k-1)\textexclamdown}{\frac{1}{(M^{2})^{k}}}e^{{-s/M^{2}}} 
%\end{eqnarray}
%The Borel transforms of some useful functions are obtained as
%\appendix*
%\newpage

%
%
\end{document}